\documentclass[12pt,preprint]{aastex}

\shorttitle{Turbulence in Supernova Remnants}
\shortauthors{Spitler and Spangler}

\begin{document}

\title{Limits on Enhanced Radio Wave Scattering by Supernova Remnants}

\author{Laura G. Spitler and Steven R. Spangler}
\affil{Department of Physics and Astronomy, University of Iowa, Iowa City, IA 52242}

\begin{abstract}
We report multifrequency observations with the NRAO Very Long Baseline Array (VLBA) of the compact radio sources J0128+6306 and J0547+2721,  which are viewed through the supernova remnants G127.1+0.5 and S147, respectively.  Observations were made at frequencies of 1.427, 1.667, 2.271, and 4.987 GHz.  The lines of sight to these sources pass through the shock wave and upstream and downstream turbulent layers of their respective supernova remnants, and thus might detect cosmic-ray generated turbulence produced during the Fermi acceleration process.  For both sources, we detect interstellar scattering, characterized by a component of the angular size which scales as the square of the observing wavelength.  The magnitude of the scattering is characterized by an effective scattering angular size $\theta_{S0}$ at a frequency of 1 GHz of $13.2 \pm 2.6$ milliarcseconds (mas) for J0128+6306 and $6.7 \pm 2.2$ mas  for J0547+2721. These angular sizes are  consistent with the ``incidental'' scattering for any line of sight out of the galaxy at similar galactic latitudes and longitudes. There is therefore no evidence for enhanced turbulence at these supernova remnants.  We establish upper limits to the supernova remnant-associated scattering measures of $8.1-14.8$ m$^{-20/3}$-pc for J0128+6306 and $3.0$ m$^{-20/3}$-pc for J0547+2721. 
\end{abstract}


\keywords{supernova remnants---cosmic rays---plasmas---turbulence---waves}

\section{Introduction}
It is widely believed that the cosmic rays are accelerated by supernova remnants (SNR) via the Fermi Class I mechanism.  The arguments in favor of this viewpoint are (1) the Fermi class I mechanism has been demonstrated to be an effective acceleration mechanism at shocks in the solar wind (e.g. \cite{Lee82}), (2) supernova explosions are one of the few obvious phenomena which generate sufficient power to account for the cosmic rays, and (3) the detection of synchrotron x-rays associated with the supernova remnant SN 1006 indicates electron acceleration to extremely high energies (\cite{Dyer01,Dyer04}).  This last observation suggests that ions are accelerated to very high energies as well.  
Nonetheless,  direct, or at least entirely convincing demonstration of ion acceleration to energies of $\sim 10^{15}$ eV or greater is not available. However, observational evidence for ion acceleration in supernova remnants has recently been considerably strengthened by detection of TeV $\gamma$-ray emission from the SNR G347.3-0.5 (\cite{Aharonian04}).  

The present paper deals with a radioastronomical method of searching for an indicator of ion acceleration at supernova remnants. The method is motivated by observed properties of shock waves in the interplanetary medium.  In the solar wind,  ion acceleration at shock waves such as the Earth's bow shock or traveling interplanetary shocks is closely connected with the generation of large amplitude magnetohydrodynamic waves (\cite{Lee82}).  Examples in the case of the Earth's bow shock are shown in \cite{Spangler88}, which also provides references to the extensive literature.  These waves are generated by cosmic rays streaming away from the shock, and constitute a ``foreshock'' upstream of the shock proper.  These waves serve to isotropize the particles without reducing their energy, so they may be repeatedly overtaken by the shock and thus accelerated.  

These waves also generate substantial density fluctuations (\cite{Spangler88,Spangler97}), which will modify the refractive index for radio waves. In the case of the Earth's foreshock, these waves are generated on a scale defined by the cyclotron resonance condition for the reflected ions that generate these waves.  Since these reflected ions have speeds many times the Alfv\'{e}n speed, the excited waves have wavelengths much greater than the ion inertial length $V_A/\Omega_i$, where $V_A$ is the Alfv\'{e}n speed and $\Omega_i$ is the ion cyclotron frequency.  \cite{Spangler97} report that the dominant wavelength for the upstream waves is typically 80-160 ion inertial lengths.  

Radio waves propagating through a region of shock-associated plasma waves will be scattered, producing detectable radio scintillation phenomena, if the region is extensive enough.  \cite{Woo85} and \cite{Woo91} reported enhanced radio wave scattering associated with interplanetary shock waves.  These papers describe numerous cases in which strong scattering transient events appeared on lines of sight passing close to the Sun.  It is possible to attribute these transients to interplanetary shock waves because the transients were preceded by a solar flare or coronal mass ejection (CME), and followed by direct detection of a shock in interplanetary space by a spacecraft such as Helios.  It is not clear from these observations how much of the enhanced scattering is due to turbulence in the foreshock, and how much is due to the downstream shocked plasma or the driver gas associated with the CME.  All of these regions have counterparts in supernova remnants.  

Radio scintillation observations further out in the interplanetary medium, at heliocentric distances $\sim 0.5$ a.u., and typically made at longer radio wavelengths, also show scattering transients. These latter transients are usually associated with corotating interaction regions (CIR) rather than shock waves (\cite{Ananth80}).  However, during particularly large flares or CMEs, transients can be observed at relatively long wavelengths (e.g. 90cm) and at heliocentric distances of several tenths of an a.u.  An example is the large transient associated with the Bastille Day flare and halo CME in July 2000 (\cite{Tokumaru03}).  Obviously, the most powerful interplanetary shocks provide the best analogs for supernova remnant shock waves.   

The idea of looking for phenomena similar to those observed in heliospheric shocks at SNR shocks was the basis of radioastronomical observations carried out and analysed nearly two decades ago by \cite{Spangler86} (Paper I) and \cite{Spangler87} (Paper II).  In those projects,  a very specific prediction of Fermi class I models was examined, i.e. that intense turbulence, and thus enhanced radio wave scattering, would be found in the foreshock ahead of the main shock jump identified in x-ray or radio continuum data.  The idea is illustrated in Figure 1, which is taken from Paper II and adapted as described below. 

In Papers I and II, sources were found whose lines of sight passed close to, but upstream of the supernova shock as identified on the basis of radio synchrotron radiation.  Thus the line of sight would not have passed through the downstream shocked region, but would have passed through the turbulent foreshock.  The idea of these investigations was that detection of enhanced radio wave scattering in this foreshock would have been a strong indicator of enhanced waves or turbulence in this region,  and thus support for the Fermi class I mechanism.  

No convincing cases were found of enhanced scattering which could be attributed to enhanced turbulence in a SNR foreshock.  On the contrary, a number of cases were found in which a line of sight to an extragalactic radio source passed very close to a supernova shock, and yet the source showed a very low level of radio wave scattering in the form of angular broadening.  The best examples were the source 3C418 near the supernova remnant HB21 (Paper I), the source 0503+467 near the remnant HB9 (Papers I, II), and, to a lesser extent, the source 0016+731 near CTA 1 (Paper I).  The most likely explanation for the failure to detect enhanced radio wave scattering is that the radio lines of sight,  although close on the sky to the SNR shells, passed upstream of the turbulent foreshock regions. In terms of the parameters defined in Figure 1, this would require the foreshock thickness $\Delta R \leq y$ where $y$ is the closest approach of the line of sight to the SNR shock.  The plausibility of this suggestion is discussed in Papers I and II. 

In the present paper, a modification of the approach of Papers I and II is employed, which circumvents the possibility that the line of sight could pass upstream of the turbulent foreshock.  We now observe radio sources viewed through supernova remnants, with the line of sight indicated by the dotted line in Figure 1.  The advantages of observing a radio source through a supernova remnant are twofold.  First of all, as is obvious from the cartoon,  the line of sight must pass through the foreshock,  so the radio measurement is affected by this region at some level.  Second, the mean plasma density is higher in the downstream region by the shock compression factor, so there is more plasma to fluctuate.  Finally, theoretical arguments indicate that shock waves should act as amplifiers of MHD turbulence, in the sense that the magnitude of fluctuations in magnetic field and plasma velocity will increase at the shock discontinuity (\cite{McKenzie68,McKenzie69,Zhuang82,Vainio98}). Since these magnetic and velocity fluctuations are the drivers of the density fluctuations,  the density fluctuations responsible for radio wave scattering effects should be magnified at the shock as well.  Thus, one might expect the SNR shock  to increase the detectability of the turbulence which is upstream of the shock. 

The obvious disadvantage to downstream observations is that detection of enhanced scattering is ambiguous,  in that it could be due to turbulence generation mechanisms in the post-shock fluid which are unrelated to plasma processes in cosmic ray acceleration.  Nonetheless, any detection of enhanced radio scattering in the vicinity of a SNR shock would be of interest, and the magnitude of the effect can be considered an upper limit to turbulence in the foreshock. 

The specific sources which have been chosen for observation are J0128+6306, an extragalactic radio source which appears nearly dead-center in the supernova remnant G127.1+0.5, and J0547+2721, which is viewed through the limb of the radiative supernova remnant S147.  More description of the sources and associated remnants is given in Section 3 below.

It might be thought that the enhanced scattering due to a small, highly turbulent  region in a supernova remnant, characterized by a size scale of a few parsecs, could not produce a detectable enhancement of the general interstellar scattering which is due to the effect of lines of sight of several kiloparsecs in length.  This point is addressed quantitatively in Section 5 below.  However, an immediate comment is that HII regions and interstellar bubbles associated with star formation regions, which have similar spatial extents, are very strong scatterers of radio waves and dominate the ``incidental'' scattering of the interstellar medium (\cite{Cordes85,Spangler98}). It is therefore plausible that highly atypical conditions near a SNR shock, such as high levels of wave turbulence, could produce interstellar scattering that dominates that of the rest of the interstellar medium.    

The organization of this paper is as follows. Section 2 briefly summarizes the relevant aspects of interstellar scattering for this project.  Section 3 describes the details of the observations and the way in which the data were processed to yield the information of interest.  Section 4 yields the observational results, including our estimates of the degree of scattering in the two supernova remnants.  Section 5 interprets those results in terms of the properties of turbulence near the SNR shocks of our two remnants.  Finally, Section 6 gives the conclusions.

\section{Relevant Formulas for Interstellar Radio Scintillation}
In this paper, we will employ the same techniques and observables as in Papers I and II.  The purpose of this brief section is to  introduce the vocabulary and formulas that will be used later, and to clarify the sort of measurements which are made.  

Interstellar turbulence possesses plasma density fluctuations, which subsequently produce random radio wave propagation (\cite{Rickett90}).  These density fluctuation are described in terms of their spatial power spectrum, $P_{\delta n}(q)$, where $q$ is the spatial wavenumber of the fluctuation.  The standard form assumed for this power spectrum is 
\begin{equation}
P_{\delta n}(q) = C_N^2 q^{-\alpha}
\end{equation}
where $C_N^2$ is the normalization constant of the density power spectrum, and will be the primary physical quantity of interest in this paper. In intuitive terms it may be thought of as the ``intensity'' of the turbulence. The spectral index $\alpha$ has been shown to be close to the Kolmogorov value of $\alpha = 11/3$ for many environments in the interstellar medium, and that value will be assumed in this investigation. 

Radio waves propagating through the interstellar medium undergo a host  of random propagation effects, such as intensity scintillations, ``corrugation'' of the spectrum, and angular broadening or blurring.  This paper concentrates on measurements of angular broadening,  which is a phenomenon which can be measured with Very Long Baseline Interferometers.  We will measure the excess angular size due to scattering in the interstellar medium.  

A point source viewed through a turbulent plasma with a power law spectrum of density fluctuations is not precisely a Gaussian image, but the difference between a best-fitting Gaussian and the true intensity distribution is relatively subtle and difficult to measure (e.g. \cite{Spangler88b}).  It has been demonstrated that the best-fit Gaussian angular size can yield good quantitative measures of turbulent properties (see \cite{Cordes85}). This matter is discussed further in Section 5.3 below.   

Radio propagation measurements yield a value for the {\em scattering measure}, which is defined as 
\begin{equation}
SM = \int_{LOS} C_N^2 ds \equiv C_N^2 Z_{eff}
\end{equation} 
where the integral is taken to be along the line of sight. In the second expression, $Z_{eff}$ is the effective thickness of a turbulent layer causing the radio propagation effects.  

Extragalactic radio sources, which are used as the probes in the present investigation as well as those of Papers I and II and many other investigations, are not point sources, and it is not always the case that the intrinsic size is much smaller than the angular broadening size.  In this case, the measured structure of the radio source is the convolution of the intrinsic structure with the angular power pattern due to interstellar scattering. When the intrinsic structure is barely resolved by the interferometer, and can be modeled as a Gaussian, the Gaussian-equivalent angular size of the source as measured is the quadratic sum of the intrinsic size ($\theta_I$) and scattering angular size ($\theta_S$).  

The intrinsic and scattered components of the angular size can be separated via multifrequency observations, because they have different dependences on the frequency of observation.  The scattering size scales as very close to the inverse square of the radio frequency; this is a consequence of the exact wavelength-squared dependence of the phase structure function of a scattered source. The inverse-frequency-squared dependence of the effective angular size has been verified for well-observed and analysed sources such as 2013+370 (\cite{Spangler88b}).  The intrinsic size, on the other hand, has a much less pronounced dependence on the frequency of observation.  For the typical case of a self-absorbed synchrotrion source with a flat spectrum, the angular size is approximately inversely proportional to the frequency (\cite{Marscher77}), a theoretical result which is borne out by extensive observational experience.  

A model expression for the measured angular size of a source which is affected by both interstellar scattering and intrinsic structure is (\cite{Fey89})
\begin{equation}
\theta_{obs}^2(\nu) = \frac{\theta_{I0}^2}{\nu^2} + \frac{\theta_{S0}^2}{\nu^4}
\end{equation}
where $\theta_{obs}(\nu)$ is the measured Gaussian equivalent size at a frequency  $\nu$ (in GHz), $\theta_{I0}$ is the intrinsic angular size at a frequency of 1 GHz, and $\theta_{S0}$ is the scattered angular size at 1 GHz. Multifrequency observations can solve or fit for the parameters $\theta_{I0}$ and $\theta_{S0}$. By convention, all angular sizes are taken to be full width at half maximum (FWHM) of the brightness distribution.   

Once the scattering size  $\theta_{S0}$ is known, the following formula can be used to obtain the scattering measure (equation (2) of Paper I, obtained from \cite{Cordes85})
\begin{equation}
 \theta_{S0}= 2.24 \left( C_N^2 Z_{eff}\right)^{3/5}  \mbox{ mas}
\end{equation}
where $Z_{eff}$ is in parsecs.  

The analysis employed in this paper uses the concepts and formulas presented above.  We acknowledge that this is far from the most sophisticated analysis which can be done with radio scattering measurements.  However, the sort of observations discussed here, which were successfully employed in Papers I and II and \cite{Fey89}, can be considered as analogous to  a litmus paper test, i.e. a relatively simple measurement that can test for the presence of interstellar scattering, and give a good quantitative measure of its magnitude as described by the scattering measure.  Values of the scattering measure obtained in this way compare very well with subsequent observations utilizing more sophisticated analyses of a larger amount of data (\cite{Fey89}).    

\section{Observations and Data Reduction}
\subsection{Sources Observed}
Following the negative results of Papers I and II, a literature search was made for supernova remnants with strong, compact radio sources in the background, so the line of sight to the compact source would have the geometry shown in Figure 1.  This search yielded two particularly good cases in which a relatively strong, compact source,  suitable for VLBI observations, was viewed through a supernova remnant.  The first is J0128+6306, which is seen through the supernova remnant G127.1+0.5, and the other is J0547+2721, viewed through the supernova remnant S147. 
\paragraph{J0128+6306/G127.1+0.5} A radio continuum image of the supernova remnant G127.1+0.5 is given  in \cite{Joncas89}, who provide a current discussion of this SNR.  The 1420 MHz continuum image of Joncas et al also clearly shows a bright radio source, nearly exactly in the center of this remnant.  This is J0128+6306. Although once proposed  as  possible SS433-type object, J0128+6306 has been firmly established as an extragalactic source through an optical ID and via HI absorption measurements (\cite{Kaplan04}).  \cite{Kaplan04} give a distance of 1.3 kpc for G127.1+0.5, based on an association with the open star cluster NGC 559. The supernova remnant has apparently not been detected in x-rays, so information is not available on the shock speed and density of the upstream interstellar medium.    
\paragraph{J0547+2721/S147} The supernova remnant S147 is a prominent optical SNR.  The radio structure of this remnant is discussed in \cite{Kundu80}.  \cite{Kaplan04} give a distance of 1.2 kpc, but other estimates range from 800 parsecs to 1.8 kpc.  An argument in favor of the closer distance is given by the observations of \cite{Salmen04}, who observed interstellar absorption lines in the spectra of distant stars.  They observed high velocity absorption lines, obviously associated with the supernova remnant, in stars with distances of 880 and 1800 parsecs.  The observations of \cite{Salmen04} show relatively low ionization state absorption lines out to velocities of $\sim \pm 70$ km/sec. This observation indicates a relatively slow shock wave with a speed $\leq$ 100 km/sec, which is consistent with other observations cited by \cite{Salmen04}, as well as the absence of x-ray emission (\cite{Sauvageot90}).  All the evidence indicates that S147 is an old, slow, radiative supernova remnant.  The compact radio source J0547+273 is located at the extreme edge of S147, just inside a faint outer filament, so the line of sight does pass through the interior of the remnant.  

\subsection{Observations} Observations of the radio sources J0128+6306 and J0547+2721 were made with the Very Long Baseline Array (VLBA) of the National Radio Astronomy Observatory\footnote{The National Radio Astronomy Observatory is managed by Associated Universities Incorporated under contract from the National Science Foundation} on 6 October 2002. Observations were also made of the strong compact radio source J0555+3948 to monitor array performance and guarantee quality of the calibration.  

Observations were made of all three sources at frequencies of 4.987, 2.271, 1.667, and 1.427 GHz.  The intermediate frequency bandwidth was 64 MHz at all frequencies except 1.427 GHz, where the need to avoid radiofrequency interference limited the recorded bandwidth to 32 MHz.  The observational protocol was to observe each of the three sources for ten minutes at a given frequency, then tune to the next frequency and repeat a set of ten minute scans.  The duration of the observing session was slightly over 10 hours.  

The data tapes were correlated with the VLBA correlator in Socorro, New Mexico.  The correlated data were processed in the AIPS software package.  The program CALIB was used to fringe-fit the data for fringe rate and delay.  Subsequent to removal of the fringe rate and delay,  the visibility data were averaged over time and frequency to produce one complex visibility measurement per baseline per 30 second time interval.  The amplitudes were calibrated by use of measured system temperatures and known antenna sensitivities.  

More limited, but nonetheless useful observations of J0128+6306 and J0547+2721 were made with a Mark III VLBI interferometer in 1988, at frequencies of 4.99 and 1.65 GHz.  A description of the technical details of those observations is given in Section 3.2 of \cite{Spangler98},  which used data from the same observing session, and thus is not repeated here.  

\subsection{Data Analysis} The primary measurement was a fit to the data in the (u,v) plane for a Gaussian equivalent angular size of the source.  The correct interpretation of this measurement requires that it refer to a compact, structurally simple source.  The ideal situation is one in which the intrinsic size is unmeasurably small, and all structure is due to interstellar scattering.  To insure that the analysis was undertaken on isolated, structurally-simple sources, we mapped and cleaned all three sources.  The maps of J0547+2721 showed simple, barely-resolved structure at all frequencies, so we could proceed directly with the analysis described in Section 2. 

The source J0128+6306 showed evidence of extended structure at the lower two frequencies.  This is indicated in Figure 2, where we show the maps at (a) 1.427 GHz  and (b) 4.987 GHz.  At the higher frequency, the source is simple and barely resolved. At 1.427 GHz, there is  extended emission about the compact component. This is very common with such sources, and the extended emission is more conspicuous at lower frequencies due to its steeper spectrum.  The effect of this emission needs to be removed before an analysis employing equations (3) and (4) can be done.  

We therefore used the AIPS task UVSUB to subtract the CLEAN components corresponding to the extended structure shown in Figure 2 from the 1.427, 1.667 and 2.271 GHz visibility data for J0128+6306.  This corrected (u,v) data set was then fit for an effective angular size $\theta_{obs}$. 

The calibrated (u,v) data (corrected, if need be for extended emission) were fit for the effective Gaussian angular size in three ways. First, we wrote a routine in {\em Mathcad} to produce a least squares fit to the visibility amplitude $V(u,v)=S_c(u,v)/S_0$ where $S_c(u,v)$ is the correlated flux density  and $S_0$ is the total source flux, to yield a value of $\theta_{obs}$.  We also used the AIPS task UVFIT, and operated it in two ways.  In the first, the total flux density $S_0$ (or equivalently, the zero spacing flux) was fixed at its measured value.  In the second approach, $S_0$ was also treated as a parameter to be fit, so it could be adjusted within a small range.  The full $(u,v)$ range of the measurements was not included in these fits. Baselines were not included for which the source was significantly resolved (e.g. a visibility less than $\sim 0.20$), since for such high degrees of resolution, a Gaussian  is not a good approximation to the visibility function (see Section 5.3).  

The three methods yielded results in very good agreement.  The maximum discrepancy of an estimate from the mean of the three estimates is 10 \%, which we adopt as the error associated with our measurements.  An illustration of one of our (u,v) data sets and the best-fit Gaussian visibility function is shown in Figure 3 as the solid line. The dashed and dotted lines are discussed in Section 5.3 below. 

The 1988 data were also fit in the $(u,v)$ plane for an effective angular size.  In what follows, we have only utilized the 1.65 GHz measurements from the 1988 session, since the lower frequency gives information on the scattering contribution to the angular size. 

\section{Observational Results}
Equivalent Gaussian angular sizes were obtained at all four frequencies for the sources 
J0128+6306, J0547+2721, and J0555+5948. As expected, the effective angular size of the calibrator source J0555+5948 was substantially smaller at all frequencies of observation. The measured angular sizes for J0128+6306 and J0547+2721 are given in Table 1. The angular sizes given in Table 1 were then fit to equation (3), yielding values for the parameters $\theta_{S0}$ and $\theta_{I0}$.  Errors on $\theta_{S0}$ and $\theta_{I0}$ were determined by the ranges in each quantity which were consistent with an acceptable reduced $\chi^2_{\nu}$.  The limit which was chosen for the acceptable range was $\chi^2_{\nu} \leq 2.60$, corresponding to a likelihood of chance occurrence of 5 \% for a fit with $\nu=3$ degrees of freedom. The values quoted in Table 1 and used throughout the paper are confident at the 95\% level.   

Values for $\theta_{S0}$ and $\theta_{I0}$ for the sources J0128+6306 and J0547+2721 are given in Table 2.  The content of the fourth through sixth columns will be described below.  The model equation (3) provided a good fit to the data for both sources.  This is illustrated in Figure 4, which shows for each source equation (3) with the angular size parameters given in Table 2.  The fact that equation (3) provides a good fit to the measured angular sizes at the four frequencies with the angular size parameters given in Table 2 supports our detection of  angular broadening due to interstellar turbulence along the line of sight to each of the radio sources.  

Interstellar scattering at a measurable level is expected along these low-latitude lines of sight,  so we next consider how much of the angular broadening illustrated in Figure 4 and parameterized in Table 2 is due to the supernova remnants, and how much is ``incidental'' angular broadening expected for such  low-latitude lines of sight, and which would be present even in the absence of a supernova remnant.  The incidental scattering may be estimated from \cite{Lazio98}, who used a large body of pulsar scintillation data to assemble a model of galactic radio wave scattering.  \cite{Lazio98} reported their estimates of  angular broadening in terms of our parameter $\theta_{S0}$.  The published table in Lazio and Cordes (1998) was supplemented by an updated version of the model available on the internet.
The published table in \cite{Lazio98} had a value for J0547+2721, but not for J0128+6306. Data are available for both sources in the updated table.  The expected values of incidental interstellar scattering provided by the Lazio and Cordes model are given by the parameter $\theta_{inc}$ (column 4 in Table 2).  

Our conclusion from Table 2 is that the measured angular broadening ($\theta_{S0}$) for our two sources is consistent with the expected, incidental broadening for any line of sight through this part of the galaxy ($\theta_{inc}$). Our measured value of $\theta_{S0}$ for J0128+6306 is marginally larger than $\theta_{inc}$,  and in fact the nominal value for $\theta_{inc}$ is outside (smaller than) the allowed range of $\theta_{S0}$ at the 95 \% confidence level.  This small excess could justify additional investigation of this source-supernova combination,  such as by measuring $\theta_{S0}$ for sources near, but outside the SNR G127.1+0.5.  Nonetheless,  our value for $\theta_{S0}$ only slightly exceeds $\theta_{inc}$ ($< 40$ \%),  and there is overlap between the statistically-allowed ranges of the two parameters.  For the remainder of this paper,  we will use our measurement of $\theta_{S0}$ as an upper limit to scattering attributable to the supernova remnant.  For the case of J0547+2721, the measured interstellar scattering is completely consistent with the a-priori estimate of incidental scattering.  In fact, the nominal value of $\theta_{S0}$ is actually less than the nominal value of $\theta_{inc}$.  

Another property of radio scattering is anisotropy of the broadened image, a characteristic which contains information on the anisotropy of the density irregularities and their distribution along the line of sight. Image anisotropy, characterized by the axial ratio ($AR$) of the scattered image and the orientation of the major axis of the image, is best measured for highly scattered sources whose structure is dominated by interstellar scattering.  This condition of heavy scattering is not well satisfied for our sources, particularly J0547+2721, so an extensive investigation of anisotropy is not made here.  

Nonetheless, we did briefly examine our data for evidence of large image anisotropy which might provide qualitative evidence of a localized, anomalous scattering layer.  The fits to the visibility data described in Section 3.3 returned separate estimates for the major and minor axes of the Gaussian brightness distribution (together with errors on these quantities).  The major and minor axes were combined in the preceding analysis to yield $\theta_{obs}$.  The results of these fits at 1.427 and 1.667 GHz were examined for evidence of scattering anisotropy. At these two lower frequencies the scattering size is maximized relative to the intrinsic size.  For J0547+2721 $AR = 1.4 \pm 0.2$ at both 1.427 and 1.667 GHz.  Although this value is comparable to those reported for sources behind the Cygnus OB1 association (\cite{Spangler98}), it may also reflect asymmetry of the intrinsic structure.  For the more heavily scattered source J0128+6306, the axial ratios are $1.20 \pm 0.14$ at 1.427 GHz and $1.04 \pm 0.41$ at 1.67 GHz.  

The main conclusion to be drawn from these numbers is that while anisotropy factors between 1.0 and 1.5 might be present for these sources,  the data are also consistent with isotropic scattering (particularly for J0128+6306).  In any case there is no evidence for highly anisotropic scattering as might happen if the scattering were dominated by a small region with anomalous turbulence along the line of sight.    

In summary, our basic observational result is that there is no significant  enhancement of the radio wave scattering to these sources due to the supernova remnants.  In the next section, we provide a quantitative analysis of this result.    

\section{Interpretation of Results}
In this section, we use our data to limit the properties of cosmic-ray-generated turbulence near supernova remnants. The data in Table 2 can limit the scattering measure associated with these two supernova remnants.  We make use of formula (4) which relates angular broadening to the physical parameter of interest, the scattering measure.  

The scattering measure to a line of sight of interest, containing an object such as a supernova remnant which contributes an excess scattering measure, can be written as $SM = SM_i + \delta SM$. The variable $SM_i$ is the incidental scattering measure for that line of sight, which would be measured in the absence of the supernova remnant, and $\delta SM$ is the excess scattering measure due to the supernova remnant.  Using equation (4), it is straightforward to derive the following formula, 
\begin{equation}
\frac{\delta SM}{SM_i} = \left(\frac{\theta_{S0}}{\theta_{inc}} \right)^{5/3} - 1
\end{equation}
 We can calculate $SM_i$ using the estimates of $\theta_{inc}$ and equation (4).  Since we are reporting all our results as upper limits,  we likewise will report an upper limit on $\delta SM$.   

\subsection{J0128+6306} The most secure upper limit to the interstellar scattering of J0128+6306 is expressed by the upper limit to $\theta_{S0}$ at the 95 \% confidence level, $\theta_{S0} \leq 15.8$ mas.  Use of this value and the value for $\theta_{inc}$ given in column 4 of Table 2 gives 
$\frac{\delta SM}{SM} \leq 1.34$. A value of  $\theta_{inc} = 9.5$ mas corresponds to an incidental scattering measure of 11.1 m$^{-20/3}$-pc.  These numbers then yield an upper limit to the supernova-remnant-associated scattering measure of $\delta SM \leq 14.8$ m$^{-20/3}$-pc.  This value is given in the fifth column of Table 2.  

A smaller, and perhaps more realistic upper limit on $\delta SM$ (although not the largest consistent with the data) is given by using the nominal value of $\theta_{S0}$ (i.e. the best-fit value) in equation (5), rather than the upper limit at the 95 \% confidence level. This value is 13.2 mas.  When this is done, we obtain $\delta SM \leq 8.1$  m$^{-20/3}$-pc.  This value is listed in parentheses in the fifth column of Table 2.  

\subsection{J0547+2721} For this source, the nominal or best-fit value of $\theta_{S0}$ actually falls below the estimated value of $\theta_{inc}$.  This obviously reflects the estimation error associated with both the  $\theta_{inc}$ and  $\theta_{S0}$ values. For purposes of extracting  an upper limit to $\delta SM$ from these data, we took $\theta_{S0}$ to be the upper limit at the 95 \% confidence level.  This gave $\theta_{S0} \leq 8.9$ milliarcseconds, and yields a corresponding value of $\delta SM \leq 3.0$ m$^{-20/3}$-pc.  This value is also given in column 5 of Table 2.  

\subsection{Model Visibility Functions with Measured $SM$} 
With these estimates of the scattering measures ($SM$ as well as $\delta SM$) for both sources, obtained via the methods discussed in Section 2, we can compare our data with a more accurate model for the visibility functions of these sources. We consider a source which possesses intrinsic, resolvable structure that can be approximated with a Gaussian brightness distribution.  This source is also subject to angular broadening due to Kolmogorov turbulence.  The visibility function of the source is a product of two functions, one a  Gaussian representing the intrinsic structure, and the second describing the angular broadening of a point source viewed through a medium possessing Kolmogorov turbulence.  In the analysis of Section 2, this broadening function was also approximated by a Gaussian,  but a more accurate function is given immediately below.  The composite visibility function (the measured correlated flux density normalized by the total flux of the compact component) is 
\begin{eqnarray}
V(u,v) = e^{-\frac{\pi^2}{4 \ln 2} \theta_I^2 (u^2 + v^2)} e^{-\frac{1}{2}D_{\phi}(u,v)} \\
D_{\phi}(u,v) = 8 \pi^2 r_e^2 f(\alpha) \lambda^{11/3} SM (u^2 + v^2)^{5/6}
\end{eqnarray}
The function $D_{\phi}$ in equation (7) is the phase structure function, and is straightforwardly obtained from eq (3) of \cite{Spangler98}.  In equation (6) $\theta_I$ represents the intrinsic angular size (FWHM) at the frequency of observation, $r_e$ is the classical electron radius, and $f(\alpha)$ is a constant involving gamma functions of the spectral index $\alpha$ defined in equation (1); for $\alpha=3.67$, $f(\alpha) = 1.18$.  

We compared eq (6) with our observations of both sources at 1.427 and 1.667 GHz.  At these lower frequencies the scattering was most pronounced.  The input parameters to eq (6) were the intrinsic size $\theta_I$, taken from the $\theta_{I0}$ values in Table 2 and extrapolated to 1.427 and 1.667 GHz, and the total scattering measure along the line of sight ($SM$), which may be calculated from $\theta_{S0}$ in Table 2 and eq (4).  For both sources the function (6) provided a satisfactory fit to the data at both of the lower two frequencies. This is illustrated in Figure 3.  The solid curve is the Gaussian model fit to the data for J0128+6306, described in Section 4 above.  The dashed line shows eq (6) with the best-fit value of the scattering measure.  The dotted line represents (6) with the upper limit to $SM$, obtained from the upper limit of $\theta_{S0} \leq 15.8$ mas.  

It may be seen that the more accurate visibility function given by eq (6) provides a satisfactory representation of the observations and, as expected, better describes the measurements on long baselines where the source is substantially resolved.  This exercise corroborates the values for the scattering measures $SM$ and $\delta SM$ obtained with the techniques described in Section 2.

\subsection{Comparison with Scattering in the Cygnus OB1 Association} It is instructive to compare the upper limits to the SNR-associated scattering measures just quoted with those due to the HII  region associated with the Cygnus OB1 association (\cite{Spangler98}).  The most weakly scattered lines of sight there (given in Table 2 of \cite{Spangler98}) had scattering measures of at least 140  m$^{-20/3}$-pc, almost all of which can be interpreted as $\delta SM$ due to Cygnus OB1 structure\footnote{The values of scattering measure  in Spangler and Cordes (1998) were reported in units of  m$^{-20/3}$-kpc; the numbers in Table 2 of Spangler and Cordes (1998) should be multiplied by 1000 to compare with the values in the present paper. }.  Even heavier scattering has been reported for the HII region NGC 6334A (\cite{Trotter98}).  From an observational standpoint, HII regions are more prominent scattering entities than supernova remnants. 

\subsection{Limits on the turbulence parameter $C_N^2$} The local measure of turbulence is the quantity $C_N^2$, which is the normalization constant of the density spectrum (Eq (1)).  This can be obtained from the scattering measurement if the thickness of the turbulent region (the parameter $Z_{eff}$ in equation (4)) is known.  As discussed in Section 3.1, reasonable distance estimates exist for both supernova remnants, being 1300 parsecs for J0128+6306 and 850 parsecs for J0574+2721.  The angular radii of the remnants can be measured from the radio map in \cite{Kundu80} to be 1.$^{\circ}$5 for S147, and from the map in \cite{Joncas89} for G127.1+0.5 to be 0$^{\circ}$.42.  We therefore estimate the radii of the supernova remnants to be 9.5 parsecs for G127.1+0.5 and 22 parsecs for S147.  

For the purposes of this paper, we are interested in the region downstream of the SNR shock in which the amplified turbulence from the upstream region remains at high levels.  This region is analogous to the magnetosheath region of the Earth's bowshock.  We do not know how extensive this region is.  As a rough estimate, we assume that the enhanced turbulence fills an annular shell of thickness equal to 10 \% of the radius of the SNR shock.  The total line of sight occupied by this intense ``SNR magnetosheath'' would then be 20 \% of the supernova radius.  Given the numbers above, $Z_{eff}$ would be 4.4 parsecs for S147 and 1.9 parsecs for G127.1+0.5.  Given the crudeness of our estimate of the size of this region, we do not attempt to account for variations in the length of a line of sight through the center of the remnant as opposed to one near the limb, as illustrated in Figure 1.  In the case of G127.1+0.5 this correction would, in any case, be very small since the line of sight goes nearly through the center of the remnant.  In the case of S147, the line of sight to J0547+2721 passes near the edge of the remnant, just inside some faint outer filaments.  The total path length in this case could be larger than twice the radial thickness, so $Z_{eff}$ could be larger,  and $C_N^2$ smaller, than we calculate below.  However, the precise location of the shock front in this part of the remnant is not certain, so we retain the simplest estimate of $Z_{eff} \simeq 0.20 R_{SNR}$, where $R_{SNR}$ is the radius of the supernova remnant. These calculations assume that the bulk of the radio wave scattering occurs in the post-shock, downstream region.  If turbulence in the foreshock also makes a significant contribution,  the true upper limits to $C_N^2$ would be smaller than those we calculate.  

With these admittedly rough estimates of $Z_{eff}$ we can convert the upper limits to scattering measure to upper limits on $\bar{C_N^2}$, the mean value of $C_N^2$ in the post-shock region.  These values are given in the sixth column of Table 2. 

We can again compare these values of   $\bar{C_N^2}$ for the Cygnus OB1 association,  given in Table 2 of \cite{Spangler98}.  It may be seen that the upper limits for J0128+6306 are actually comparable to the measured values for the most weakly scattered Cygnus OB1 sources, 2013+370 and 2012+383.  Although the scattering measures for the Cygnus sources are larger, so is the estimated thickness of the turbulent region associated with Cygnus OB1.  Thus our data cannot exclude the possibility that within the turbulent region of the shock in G127.1+0.5, the turbulent fluctuations in density are as strong as those in parts of the Cygnus OB1 region.  

Our upper limit for S147 is substantially smaller, $\bar{C_N^2} < 1$ m$^{-20/3}$, and smaller than for any line of sight through Cygnus OB1. 

Our upper limits to the parameter $\bar{C_N^2}$, although small compared to the measured values in the ionized shell surrounding the Cygnus OB1 association, are very large compared to the general background value characteristic of the interstellar medium.  The pervasive ``Type A'' turbulence (\cite{Cordes85}), which is associated with the Diffuse Ionized Gas (DIG) component of the interstellar medium, has a reasonably uniform value of $C_N^2 \simeq 3 \times 10^{-4}$ m$^{-11/3}$ (\cite{Cordes85}).  Thus, our observations are compatible with a factor of between 2000 (S147) to 30000 (G127.1+0.5) enhancement of the normal background turbulence level in the turbulent shells of supernova remnants.   

These limits on $C_N^2$ can be translated into constraints on the properties of turbulence in the foreshock and postshock regions of the supernova remnants G127.1+0.5 and S147, as discussed in Papers I and II.  Our observations limit a quantity which is determined by the rms density fluctuation $\sigma_n$ and the outer scale to the turbulence $l_0$ (see eq (5) of Paper I).  Figure 4 of Paper I or Figure 5 of Paper II can be used for an analysis of the present observations.  This sort of analysis is impeded by our poor knowledge of the plasma density either upstream or downstream in these remnants.  

Such an exercise was carried out with our data for J0128+6306 and J0547+2721, using the numbers in Table 2 and an equivalent of eq (5) of Paper I.  As a representative example of such a calculation, we assume that the upstream plasma number density is $1 \mbox{ cm}^{-3}$, which is increased to  $4 \mbox{ cm}^{-3}$ after passage through the shock.  We also adopt, as a representative value, an rms density fluctuation which is 18 \% of the mean density, so $\sigma_n = 0.72$.  This choice of modulation index is admittedly arbitrary, but is based on observed values in the Earth's foreshock (result cited in \cite{Spangler02}).  Our observations then place a lower limit on the outer scale of the turbulence of about 0.1 parsecs for J0128+6306.  For the more weakly scattered source J0547+2721, the limits are more interesting and restrictive; the outer scale would have to be almost as large as the assumed thickness of the turbulent shell, $Z_{eff}/2 = 2.2$ parsecs.  

Higher assumed values of $\sigma_n$ (resulting from a higher mean density or larger density modulation index) make the observational limits more interesting and physically restrictive.  Such limits could potentially argue against an anomalously elevated level of turbulence in the post shock region.  On the other hand, a lower value of $\sigma_n$ than that adopted in this calculation is less confining in terms of turbulence characteristics.  The situation could be improved by x-ray observations of these remnants which would provide knowledge of the mean density in the downstream region and the shock speed.

For the present we are left with the results of this paper, and Papers I and II.  There are now several cases of radio lines of sight which pass close to, or inside of a supernova remnant shock.  In no case is there an indication of an enhanced level of plasma turbulence at spatial scales of 200 - 10000 kilometers. Since the ion inertial length for a hydrogen plasma with a density of 4 cm$^{-3}$ (our representative value for the downstream density in a SNR shock) is $\simeq 120$ km, the irregularities to which our measurements are sensitive are from about 2 to 80 ion inertial lengths in extent. For an upstream region with a density of 1 cm$^{-3}$ the ion inertial length would be larger by a factor of two, and the range of irregularities sampled would be approximately 1 to 40 ion inertial lengths.  This range of scales is close to, if slightly smaller than that on which substantial density fluctuations are seen in the Earth's foreshock.  This realization was the motivation for our investigation.  However, if physical processes in an SNR shock generate compressive irregularities on much larger spatial scales, perhaps hundreds to thousands of ion inertial lengths, the anomalous turbulence might have escaped detection with our technique.

\section{Conclusions}
The results and conclusions of this paper are as follows.
\begin{enumerate}
\item Observations with the Very Long Baseline Array were made at frequencies of 1.427, 1.667, 2.271, and 4.987 GHz of the source J0128+6306, which is seen through the supernova remnant G127.1+05, and J0547+2721, which is seen through the remnant S147.  
\item The measured angular sizes of the compact portions of these sources at the four frequencies are in good agreement with a model in which the observed source structure is a convolution of an interstellar scattering image with the intrinsic structure of the source.  Our estimates of the interstellar scattering size $\theta_{S0}$ (FWHM) at 1 GHz is $13.2 \pm 2.6$ milliarcseconds (mas) for J0128+6306, and $6.7 \pm 2.2$ mas for J0547+2721.  
\item The measured  values are consistent with a priori estimates of the interstellar scattering due to known interstellar turbulence along these lines of sight. We therefore see no evidence for enhanced turbulence upstream or downstream of these supernova remnant shock waves.  Our upper limits on the angular broadening due to SNR-associated turbulence can be translated to constraints on turbulence at these supernova remnants, but the constraints are weak due to poorly known or unknown properties of the supernova remnants.   
\end{enumerate}

\acknowledgments
This work was supported at the University of Iowa by grants ATM99-86887 and ATM03-54782 from the National Science Foundation. LGS was also supported by the University of Iowa through the Student Internship Program.  We thank William Coles for a thorough and helpful review of the paper.  



\clearpage



\begin{figure}
\epsscale{.75}
\plotone{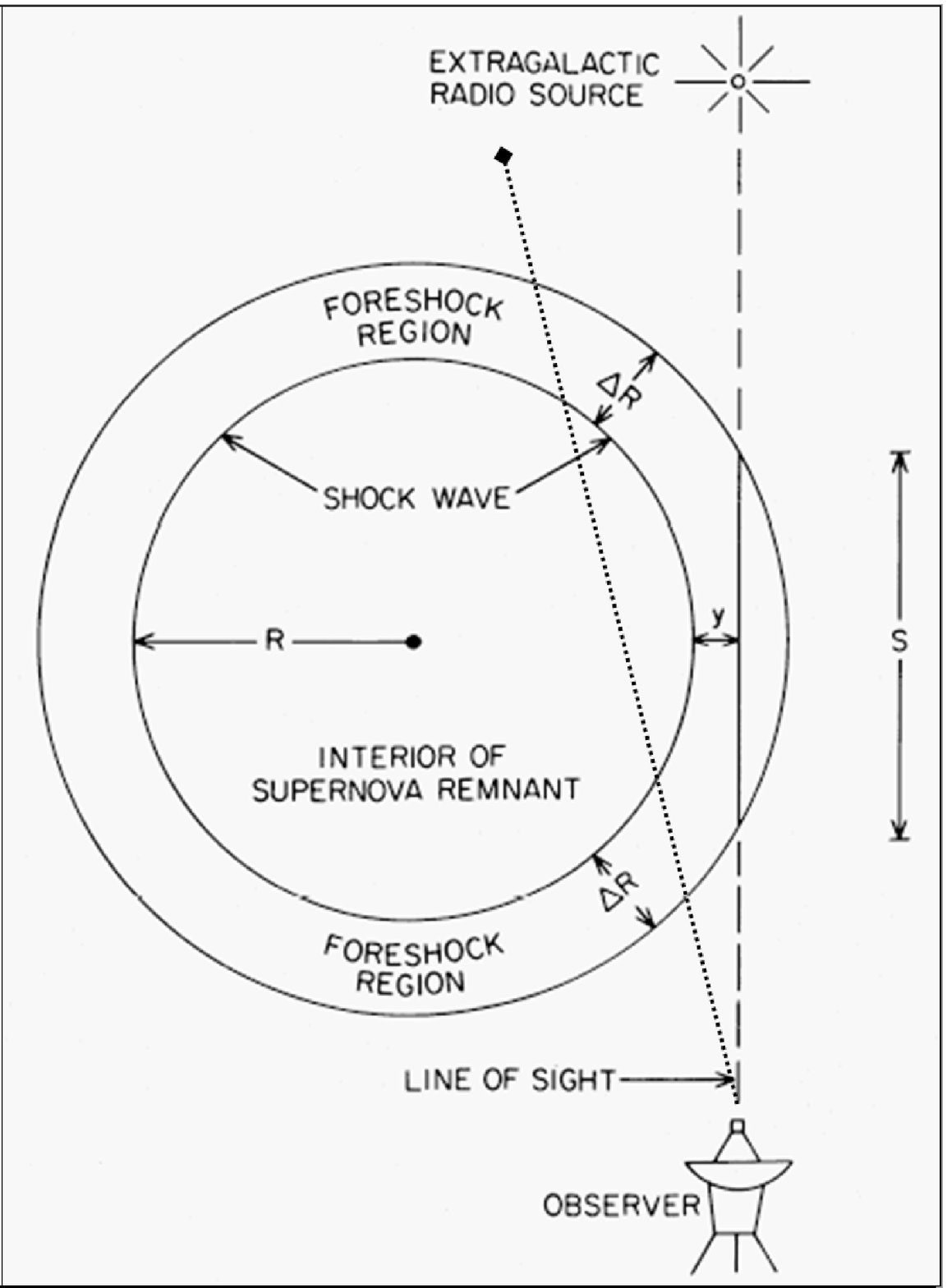}
\caption{Cartoon illustrating the idea of radio probing of the turbulent environment in the vicinity of a supernova remnant shock.  If the Fermi class I acceleration mechanism is operative,  there should be a region of intense turbulence upstream of the shock, termed the ``foreshock''.  A search for this turbulence may be made by imaging a radio source viewed through the foreshock, and see if it is angularly broadened by this turbulence.  Investigations by \cite{Spangler86} and \cite{Spangler87} showed no detectable broadening of this sort for several supernova remnants.  The present investigation deals with a more favorable situation for scattering, in which the line of sight (dotted line) passes through the interior of the remnant, where the upstream turbulence is amplified by passage through the shock. Figure adapted from \cite{Spangler87}.}
\end{figure}
\clearpage
\begin{figure}
\epsscale{.60}
\plotone{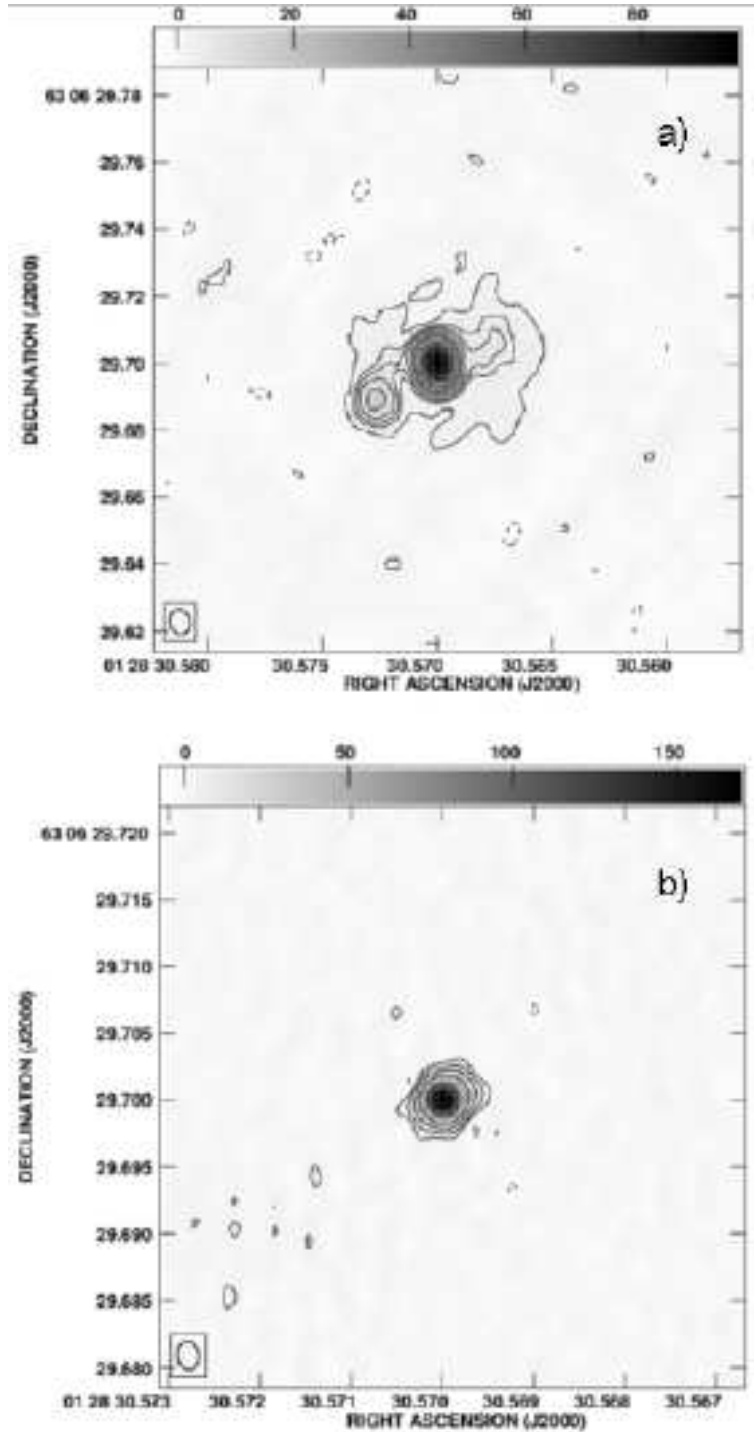}
\caption{Maps of the radio source J0128+6306 at (a) 1.427 GHz and (b) 4.987 GHz.  The lower frequency map shows the extended structure which was removed from the visibility data prior to fitting for an angular size.  The beam size is 7.61 by 5.79 milliarcseconds at 1.427 GHz and 2.15 by 1.59 mas at 4.987 GHz.  The contours in both images are at -2,2,5,10,20,30,50,70,80,and 90 percent of peak intensity in both images. The peak intensity is 0.096 Janskys per beam at 1.427 GHz and 0.168 Janskys per beam at 4.987 GHz.}
\end{figure}
\clearpage
\begin{figure}
\epsscale{.80}
\plotone{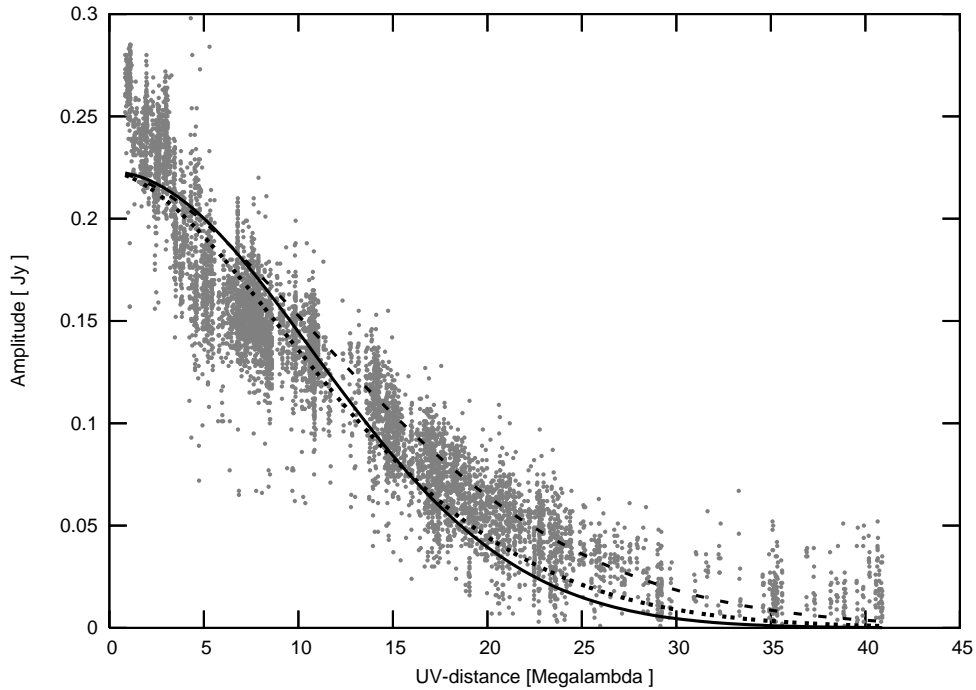}
\caption{Visibility measurements for the source J0128+6306 at a frequency of 1.427 GHz.  The solid line corresponds to the best-fit model visibility function with the Gaussian equivalent angular size in Table 1. The dashed and dotted lines show a more accurate expression for the visibility function of scattered radio source, given by eq (6).  The dashed line uses the best fit scattering measure $SM$ emergent from our analysis, and the dotted line used the upper limit at the 95 \% confidence level. }
\end{figure}
\clearpage
\begin{figure}
\epsscale{.80}
\plotone{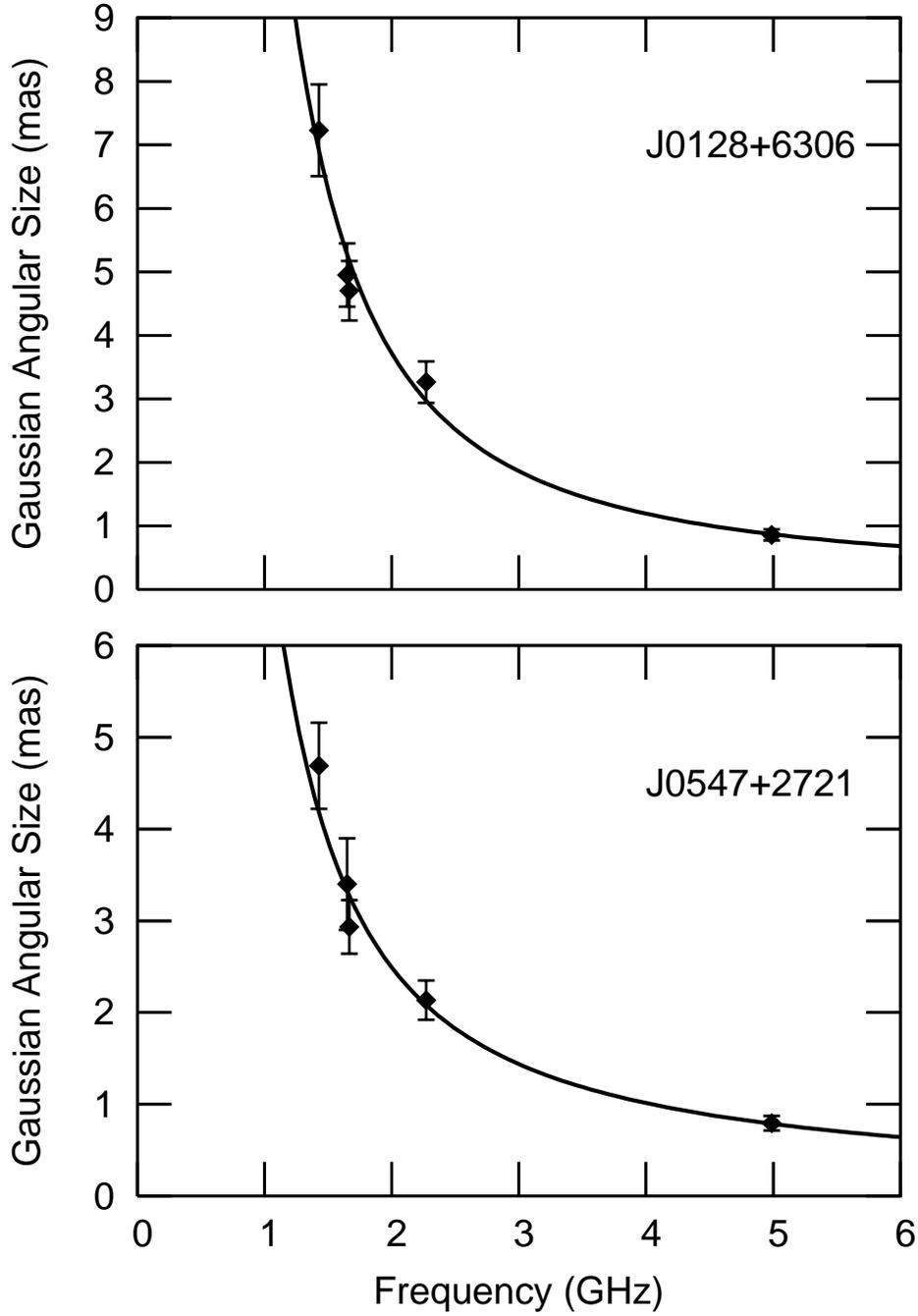}
\caption{Observed angular size as function of frequency for (a) J0128+6306 and (b) J0547+2721.  The solid curve represents the expected relation for a source whose size is a convolution of a scattering angular size with an intrinsic angular size (equation (3)) with the angular size parameters given in Table 2.   }
\end{figure}

\clearpage
\begin{deluxetable}{crl}
\tabletypesize{\scriptsize}
\tablecaption{Measured Gaussian-Equivalent Angular Sizes \label{tbl-1}}
\tablewidth{0pt}
\tablehead{\colhead{Source} & \colhead{Freq (GHz)} & \colhead{$\theta_{obs}$(FWHM,mas)}}
\startdata
J0128+6306 & 4.987 & $0.86 \pm 0.09$  \\
          & 2.271 & $3.26 \pm 0.33$  \\         
          & 1.667 & $4.70 \pm 0.47$  \\
          & 1.650  & $4.95 \pm 0.50^{\ast}$ \\
          & 1.427 & $7.23 \pm 0.72$  \\
J0547+2721 & 4.987 & $0.79 \pm 0.08$  \\
          & 2.271 & $2.13 \pm 0.21$  \\
          & 1.667 & $2.93 \pm 0.29$  \\
          & 1.650 & $3.40 \pm 0.50^{\ast}$ \\  
          & 1.427 & $4.69 \pm 0.47$  
\enddata
\tablecomments{$\ast$: Observations at 1.650 GHz made with VLBI network in 1988.}
\end{deluxetable}
\clearpage
\begin{deluxetable}{crrrrl}
\tabletypesize{\scriptsize}
\tablecaption{Retrieved Scattered and Intrinsic Angular Sizes \label{tbl-2}}
\tablewidth{0pt}
\tablehead{\colhead{Source} & \colhead{$\theta_{I0}$ } & \colhead{$\theta_{S0}$} & \colhead{$\theta_{inc}$} & \colhead{$\delta SM$ } & \colhead{$\bar{C_N^2}$}}
\startdata
J0128+6306 & $3.4 \pm 1.0$ & $13.2 \pm 2.6$ & $9.5 \pm 1.9$ & 14.8 (8.1) & 7.8 (4.3)\\
J0547+2721 & $3.7 \pm 0.8$ & $6.7 \pm 2.2$ & $9.0 \pm 1.8$ & 3.0 & 0.68
\enddata
\tablecomments{ All angular sizes are in milliarcseconds. The units of the scattering measure $SM$ and $C_N^2$ are m$^{-20/3}$-parsecs and  m$^{-20/3}$, respectively. Dual values for J0128+6306 correspond to different ways of calculating the upper limit to SNR-associated scattering.  }
\end{deluxetable}

\end{document}